\documentclass[12pt, a4paper]{article}

\usepackage[utf8]{inputenc}
\usepackage[T1]{fontenc}
\usepackage{lmodern}
\usepackage{geometry}
\usepackage{hyperref}
\usepackage{graphicx}
\usepackage{booktabs}
\usepackage{tabularx}
\usepackage{enumitem}
\usepackage{xcolor}
\usepackage{listings}
\usepackage{caption}
\usepackage{amsmath}
\usepackage{subcaption}
\usepackage{amssymb}
\usepackage{natbib}
\usepackage{url}
\usepackage{lineno}
\usepackage{microtype}
\usepackage{tikz}
\usetikzlibrary{shapes.geometric, arrows.meta, positioning, shadows, calc}
\usepackage{silence}
\usepackage{appendix}

\graphicspath{{figures/}}

\definecolor{pedalGold}{HTML}{F59E0B}
\definecolor{pedalEmerald}{HTML}{10B981}
\definecolor{pedalDark}{HTML}{0D0D0D}
\definecolor{codeBackground}{HTML}{F5F5F5}
\definecolor{codeBorder}{HTML}{CCCCCC}

\hypersetup{
  colorlinks=true,
  linkcolor=blue!70!black,
  citecolor=pedalEmerald,
  urlcolor=blue!60!black,
  pdftitle={PEDAL: Open Infrastructure for Citable AI Prompts in STEM Education and Research},
  pdfauthor={Murat Kahveci}
}

\title{\textbf{PEDAL: Open Infrastructure for Citable AI Prompts in STEM Education and Research}}

\author{
  Murat Kahveci\thanks{Correspondence: \texttt{murat@kahveci.pw}} \\
  \textit{Kahveci Nexus Research Group} \\
  \textit{Chicago, IL, USA} \\
  \textit{ORCID: 0000-0001-5155-7755}
}

\date{May 2026}

\begin{document}
\maketitle
% \linenumbers  % disabled for arXiv preprint

% =============================================================================
% ABSTRACT
% =============================================================================
\begin{abstract}
The rapid integration of Large Language Models (LLMs) into educational practice has created an urgent need for infrastructure that treats AI prompts not as disposable instructions but as reproducible scholarly artifacts. This paper presents PEDAL (Pedagogical Evaluation, Design, \& Analysis Lab), an open-research platform that implements a three-tier \textit{Laboratory-to-Archive} pipeline for prompt engineering: (1)~an \textbf{Orchestration Layer} with AI-assisted prompt generation and a Scholarly Inject toolkit for automated metadata extraction; (2)~a \textbf{Laboratory Layer} supporting Git-style version control, LLM-as-a-Judge evaluation, and Mann--Whitney~$U$ statistical testing for champion variant identification; and (3)~a \textbf{Public Archive Layer} providing scholarly publication with per-version DOI minting via Zenodo, multi-format data exports (JSON, CSV, \LaTeX), and SEO-optimized discoverability through JSON-LD and HighWire Press metadata. PEDAL's Scholarly Sync~2 (SS2) framework attaches a 24+ field metadata envelope to every artifact, encoding Bloom's Revised Taxonomy, Webb's Depth of Knowledge, SAMR integration levels, 5E instructional phases, and NGSS standards alignment. A chemistry education exemplar demonstrates the full pipeline from Socratic inquiry scaffolding through statistical evaluation to public DOI-minted archival. We further present the NExAIE (Nexus AI \& Education) extension, which applies PEDAL's infrastructure to AI-augmented peer review through a 42-prompt evaluation matrix spanning six quality dimensions across seven publication types, introducing \textit{Radical Transparency} by publicly archiving all review rubrics with DOIs. Initial deployment data---4,684 artifact views from 1,810 unique researchers within one month of launch---indicates strong community demand for citable, quality-assured AI scaffolding in STEM education and research. The complete platform is released under CC-BY-4.0 (DOI: 10.5281/zenodo.19474709).
\end{abstract}

\noindent\textbf{Keywords:} prompt engineering; large language models; AI in education; STEM education; open science; FAIR principles; reproducibility; Bloom's taxonomy; AI-augmented peer review

% =============================================================================
% §1  INTRODUCTION
% =============================================================================
\section{Introduction}
\label{sec:intro}

The emergence of Large Language Models (LLMs) as instructional tools has fundamentally altered the landscape of educational technology \citep{kasneci2023, holmes2022}. Educators, researchers, and curriculum designers now routinely craft sophisticated prompts---structured natural-language instructions---to generate lesson plans, assessment items, feedback scaffolds, and inquiry-based activities. Yet these prompts, despite embodying significant pedagogical reasoning and iterative refinement, remain largely ephemeral: shared informally, stored without version control, and discarded after use. The scholarly community lacks infrastructure for treating prompts as the \textit{research instruments} they have become.

This gap is not merely administrative. The reproducibility crisis in computational social science \citep{baker2016} and the growing recognition that methodological transparency is essential for credible AI-education research \citep{nosek2015} demand that every component of an AI-mediated educational intervention---including the prompts that drive it---be documented, versioned, and citable. The FAIR principles (Findable, Accessible, Interoperable, Reusable) \citep{wilkinson2016} provide a framework for scientific data stewardship, but to our knowledge, no existing platform operationalizes these principles specifically for AI prompt artifacts within an educational research context.

Simultaneously, the integration of LLMs into scholarly publishing itself presents both opportunities and challenges. Peer review---the cornerstone of scientific quality assurance for over three centuries \citep{kronick1990}---is under unprecedented strain. Reviewer fatigue, declining review quality, and documented biases in gender \citep{helmer2017}, institutional prestige \citep{tomkins2017}, and geography \citep{ross2006} have eroded confidence in the system. Early explorations of AI-assisted review show promising alignment with human evaluations \citep{liang2024, checco2021}, but existing implementations suffer from prompt opacity, monolithic evaluation, and conflation of AI and human reviewer roles.

This paper presents two interconnected contributions addressing these challenges:

\begin{enumerate}
    \item \textbf{PEDAL} (Pedagogical Evaluation, Design, \& Analysis Lab)---an open-research infrastructure that transforms transient AI prompts into citable, reproducible scientific artifacts through a three-tier \textit{Laboratory-to-Archive} pipeline. PEDAL operationalizes the FAIR principles for prompt engineering by providing:
    \begin{itemize}
        \item Git-style version control with branching, immutable snapshots, and dependency graphs;
        \item A Scholarly Sync~2 (SS2) metadata framework encoding pedagogical taxonomies (Bloom's, DOK, SAMR, 5E) and NGSS standards alignment;
        \item Per-version DOI minting via Zenodo with multi-format citation export;
        \item LLM-as-a-Judge evaluation with Mann--Whitney~$U$ statistical testing for rigorous variant comparison; and
        \item A role-based access control (RBAC) collaboration system with PI-gated approval workflows.
    \end{itemize}

    \item \textbf{NExAIE} (Nexus AI \& Education)---an extension module that applies PEDAL's scholarly artifact infrastructure to AI-augmented peer review, demonstrating the platform's extensibility beyond prompt archiving. This paper presents the NExAIE architecture as an \textit{implementation exemplar}; the full specification of the 42-prompt evaluation matrix, calibration methodology, and editorial workflow is detailed in a companion paper \citep{kahveci2026nexaie}. NExAIE introduces:
    \begin{itemize}
        \item A 42-prompt evaluation matrix decomposing manuscript assessment into six orthogonal quality dimensions, each calibrated against review criteria from top-tier journals (JRST, Computers \& Education, BJET, IJSE);
        \item A \textit{Radical Transparency} paradigm in which all evaluation rubrics are publicly archived with DOIs, enabling prospective authors to stress-test manuscripts against editorial standards; and
        \item A privacy-preserving dual-view architecture employing AES-256-CBC encryption, permanent pseudocoding, and token-based editorial access.
    \end{itemize}
\end{enumerate}

The central thesis of this work is that \textit{the prompts are the methodology}. Just as a laboratory scientist documents reagent concentrations, instrument calibrations, and procedural steps, an AI-mediated educational intervention must document the exact instructions given to the model. By providing infrastructure that makes this documentation natural, versioned, and citable, PEDAL transforms prompt engineering from an ad hoc practice into a rigorous scholarly discipline.

The remainder of this paper is organized as follows. Section~\ref{sec:background} reviews related work on prompt engineering, open science in AI-education research, and peer review reform. Section~\ref{sec:architecture} details the PEDAL platform architecture across its three tiers. Section~\ref{sec:ss2} presents the Scholarly Sync~2 metadata framework and research classification grid. Section~\ref{sec:stats} describes the statistical engine and Zenodo DOI integration. Section~\ref{sec:rbac} outlines the collaborative role-based access control system. Section~\ref{sec:nexaie} presents the NExAIE AI-augmented peer review extension. Section~\ref{sec:transparency} describes the Radical Transparency model and dual-view architecture. Section~\ref{sec:data} specifies the data model and security implementation. Section~\ref{sec:discussion} discusses implications, limitations, and future work. Section~\ref{sec:conclusion} concludes.

% =============================================================================
% §2  BACKGROUND & RELATED WORK
% =============================================================================
\section{Background and Related Work}
\label{sec:background}

\subsection{Prompt Engineering as Scholarly Practice}

The term ``prompt engineering'' has evolved rapidly from informal trial-and-error experimentation to a recognized subdiscipline within AI research \citep{liu2023prompting}. White et al. introduced a catalog of reusable prompt patterns---persona, template, chain-of-thought, and metacognitive scaffolding---demonstrating that structured prompt design significantly improves LLM output quality \citep{white2023}. However, the transition from professional practice to end-user accessibility remains fraught with challenges. \citet[p.~1]{zamfirescu2023} demonstrated that non-expert users struggle systematically with prompt construction, often failing because they rely on \textbf{human-to-human instructional experiences} and expectations \citep[pp.~1, 14]{zamfirescu2023}.

Instead of adopting the rigorous methodologies seen in traditional software engineering, these users exhibit \textbf{opportunistic rather than systematic} design and testing behaviors \citep[p.~2]{zamfirescu2023}. Specifically, failure modes include a tendency to over-generalize from single observations and an inclination to use socially appropriate, polite language that lacks the technical precision required for robust model steering \citep[pp.~10, 11]{zamfirescu2023}. These findings explicitly underscore the need for documented best practices and reusable templates, validating the necessity of a structured orchestration layer to bridge the gap between intuitive human conversation and effective machine instruction \citep[p.~13]{zamfirescu2023}.

Despite this growing sophistication, prompt engineering lacks the infrastructure norms that other research methodologies take for granted. A chemist publishes reagent specifications; a psychologist publishes survey instruments; a computational scientist publishes source code. Yet an educator who develops a multi-turn scaffolding prompt---potentially representing weeks of iterative refinement across multiple LLM providers---has no standard mechanism for versioning, citing, or reproducing that work. Existing prompt-sharing platforms (GitHub repositories, Hugging Face model cards, community forums) treat prompts as interchangeable code snippets rather than as scholarly instruments whose evaluative context justifies their deployment.

PEDAL addresses this gap by applying the norms of laboratory science to prompt engineering: every prompt receives a permanent identifier (DOI), version history, pedagogical metadata, and statistical evaluation evidence before it enters the public archive.

\subsection{Open Science and the FAIR Principles}

The open science movement has established that research artifacts must be Findable, Accessible, Interoperable, and Reusable (FAIR) to support reproducibility and cumulative knowledge building \citep{wilkinson2016}. The reproducibility crisis---in which 70\% of researchers reported failing to reproduce another scientist's experiments \citep{baker2016}---has prompted sweeping reforms in data sharing, pre-registration, and methodological transparency \citep{nosek2015}.

These principles have been operationalized for datasets (Zenodo, Figshare), software (Software Heritage, JOSS), and protocols (protocols.io). However, AI prompts occupy an awkward intermediate space: they are neither raw data nor executable code, but structured natural-language instructions whose effectiveness depends on context (model version, system prompt, parameter settings, input data). No existing FAIR-compliant infrastructure addresses this artifact type within an educational research context.

PEDAL's contribution is to operationalize FAIR specifically for pedagogical prompt artifacts. Each prompt is \textit{Findable} through public slugs, JSON-LD structured data, and Google Dataset Search indexing; \textit{Accessible} through a public archive with no authentication barriers; \textit{Interoperable} through machine-readable JSON metadata exports with standardized taxonomy encodings; and \textit{Reusable} through version-controlled snapshots with DOIs, citation exports, and CC-BY-4.0 licensing.

\subsection{The Crisis of Traditional Peer Review}

The limitations of conventional peer review are well-documented. Inter-reviewer agreement is consistently low, with studies reporting $\kappa$ values between 0.17 and 0.34---barely above chance \citep{bornmann2010}. Reproducibility of reviewer decisions is poor: when the same manuscripts are re-submitted to the same journal, acceptance rates diverge substantially \citep{peters1982, rothwell2000}. The process is slow, with median review times exceeding 100 days in many fields \citep{huisman2017}, and it is expensive, with estimated costs of \$1.5--2.5 billion annually in donated reviewer labor \citep{aczel2021}.

In education research specifically, these challenges are compounded by the interdisciplinary nature of the field. Journals at the intersection of AI and education---such as Computers \& Education, the British Journal of Educational Technology (BJET), and the International Journal of Science Education (IJSE)---must evaluate manuscripts spanning computer science, cognitive science, pedagogy, and domain-specific disciplinary knowledge. Finding reviewers with expertise across all relevant dimensions is increasingly difficult \citep{zawacki2019}.

\subsection{AI-Assisted Evaluation in Scholarly Publishing}

Early applications of automated review focused on surface-level quality indicators: plagiarism detection \citep{foltynek2019}, statistical reporting verification \citep{nuijten2016}, and reference formatting checks \citep{priem2022}. More recent work has explored LLMs for substantive evaluation. Liang et al. demonstrated that GPT-4 reviews overlapped substantially with human reviews, though AI reviews were more likely to identify novel weaknesses than to validate strengths \citep{liang2024}. Checco et al. developed AI-based tools for predicting review outcomes, reporting moderate accuracy but noting significant domain-dependency \citep{checco2021}. The LLM-as-a-Judge paradigm, formalized by Zheng et al. through MT-Bench and Chatbot Arena, established that LLMs can serve as reliable evaluators when given structured rubrics and calibration instructions \citep{zheng2024}.

However, the field lacks a \textit{systematic framework} for deploying AI evaluation at production scale with verifiable rigor. Existing approaches share several deficiencies:

\begin{itemize}
    \item \textbf{Prompt opacity:} The evaluation instructions given to LLMs are not disclosed, making it impossible for the scholarly community to audit, reproduce, or improve upon the review process.
    \item \textbf{Holistic scoring:} Most AI-review systems produce a single quality score or accept/reject recommendation, offering little diagnostic value to authors seeking to improve specific aspects of their work.
    \item \textbf{Genre blindness:} Evaluation criteria are applied uniformly regardless of publication type, despite the fundamentally different expectations for research articles, review papers, technology reports, and commentaries.
    \item \textbf{Confidentiality trade-offs:} Systems that achieve transparency in their evaluation process often compromise manuscript confidentiality, while those preserving confidentiality offer no insight into their methods.
\end{itemize}

\subsection{Transparency and Reproducibility in Review}

The open science movement has increasingly scrutinized the opacity of peer review itself. Open peer review initiatives---such as those at eLife \citep{elife2022} and journals adopting the Publons framework---have demonstrated that publishing reviewer reports alongside articles can improve review quality and accountability \citep{rosshellauer2017}. However, these initiatives apply to \textit{human} reviews where the reviewer's reasoning is inherently embedded in their written feedback.

AI-augmented review introduces a new transparency challenge: the evaluation is only as rigorous as the \textit{prompt engineering} that drives it. Unlike a human reviewer whose expertise and reasoning are assumed, an LLM's evaluation is entirely determined by its instructions. This observation motivates the central design principle of the NExAIE framework: \textit{the prompts themselves are the methodology}, and like any methodology, they must be documented, versioned, peer-reviewed, and publicly accessible. The PEDAL Archive \citep{kahveci2026pedal} provides the ideal infrastructure for this requirement, treating review prompts as citable scholarly artifacts.
% =============================================================================
% §3  PEDAL PLATFORM ARCHITECTURE
% =============================================================================
\section{PEDAL Platform Architecture}
\label{sec:architecture}

\begin{figure}[ht]
\centering
\begin{tikzpicture}[
    node distance=1.2cm and 1.5cm,
    box/.style={rectangle, draw=black, thick, rounded corners, fill=gray!5, text centered, minimum width=3.5cm, minimum height=1cm, font=\small},
    layer/.style={rectangle, draw=black, dashed, thick, rounded corners, fill=gray!2, inner sep=0.5cm, font=\bfseries\small},
    arrow/.style={-{Stealth[scale=1.2]}, thick},
    title/.style={font=\bfseries\small, text centered}
]

% --- LAYERS ---
\node[layer, minimum width=4.5cm, minimum height=5.5cm] (L1) at (0,0) {};
\node[layer, minimum width=4.5cm, minimum height=5.5cm] (L2) at (5.5,0) {};
\node[layer, minimum width=4.5cm, minimum height=5.5cm] (L3) at (11,0) {};

\node[title, above=0.1cm of L1] {ORCHESTRATION LAYER};
\node[title, above=0.1cm of L2] {LABORATORY LAYER};
\node[title, above=0.1cm of L3] {ARCHIVE LAYER};

% --- COMPONENTS ---
\node[box] (C1_1) at (L1.center) [yshift=1.8cm] {Generative Architect};
\node[box] (C1_2) at (L1.center) {Scholarly Inject};
\node[box] (C1_3) at (L1.center) [yshift=-1.8cm] {Artifact Ingest};

\node[box] (C2_1) at (L2.center) [yshift=1.8cm] {Version Control};
\node[box] (C2_2) at (L2.center) {LLM-as-a-Judge};
\node[box] (C2_3) at (L2.center) [yshift=-1.8cm] {Research Lab};

\node[box] (C3_1) at (L3.center) [yshift=1.8cm] {PubliView};
\node[box] (C3_2) at (L3.center) {Zenodo DOI Minting};
\node[box] (C3_3) at (L3.center) [yshift=-1.8cm] {SEO \& Indexing};

% --- METADATA SIDECAR ---
\node[box, fill=black, text=white, minimum width=8.5cm, minimum height=0.8cm] (SS2) at (2.75,-5.0) {SCHOLARLY SYNC 2 (SS2) METADATA ENGINE};

% --- EXTENSIONS ---
\node[box, fill=gray!5, dashed, minimum width=4cm] (NEXAIE) at (11, -5.0) {NExAIE EXTENSION};

% --- FLOWS ---
\draw[arrow] (L1) -- (L2);
\draw[arrow] (L2) -- (L3);
\draw[arrow] (SS2.north -| L1.south) -- (L1.south);
\draw[arrow] (SS2.north -| L2.south) -- (L2.south);
\draw[arrow, <->] (L3.south) -- (NEXAIE.north);

\end{tikzpicture}
\caption{The PEDAL three-tier Laboratory-to-Archive architecture showing the pipeline from prompt creation to DOI-minted publication, anchored by the SS2 metadata framework.}
\label{fig:architecture}
\end{figure}
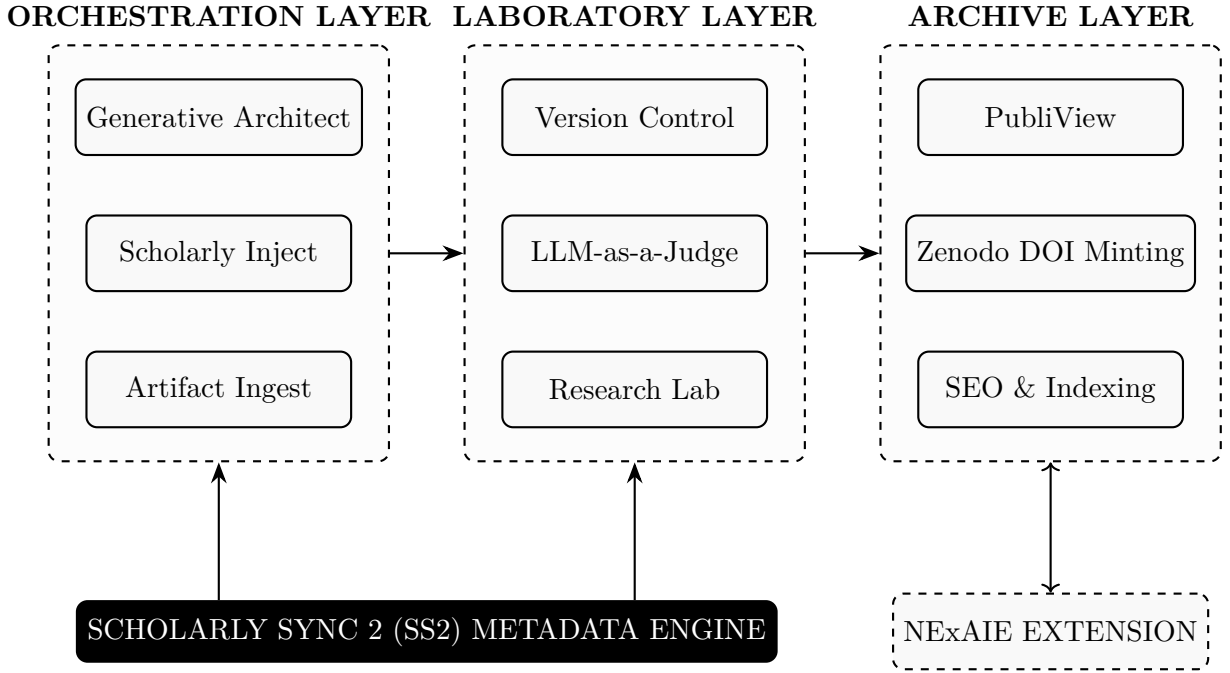

PEDAL operates as a production web application built on PHP~8.x and MySQL/MariaDB~11.8, with a vanilla JavaScript frontend using Bootstrap~5. Evolving beyond its initial ``True Dark'' industrial aesthetic, the platform now defaults to a refined light theme (\texttt{\#F4F6F8} background, \texttt{\#FFFFFF} cards). The visual language utilizes deep slate black text alongside rich gold and muted pastel accents (e.g., sage green, slate blue) to denote semantic meaning and enhance readability. Typography relies on Open Sans for headings and body text, with JetBrains Mono reserved for code and data elements. This section details the three architectural layers that implement the Laboratory-to-Archive pipeline (see Figure~\ref{fig:architecture}).

\subsection{Design Philosophy}

Three principles govern PEDAL's architecture:

\begin{enumerate}[label=\textbf{D\arabic*.}]
    \item \textbf{Prompts as Laboratory Instruments.} Every prompt is treated with the same rigor as a laboratory protocol: it must be documented, versioned, evaluated against explicit criteria, and archived with a permanent identifier before it is considered ready for public use.
    \item \textbf{Metadata-First Design.} Pedagogical context (Bloom's level, DOK, SAMR, 5E phase, NGSS alignment) is not optional annotation but structural metadata that shapes how artifacts are classified, discovered, and cited.
    \item \textbf{Separation of Orchestration, Evaluation, and Publication.} The three tiers operate independently: an artifact can be created in the Orchestration layer, refined through multiple evaluation cycles in the Laboratory layer, and published to the Archive layer only when it meets explicit quality thresholds.
\end{enumerate}

\subsection{Layer 1: Orchestration}
\label{sec:layer1}

The Orchestration layer (\texttt{dossier/\allowbreak prompt\_new.php}) provides three mechanisms for artifact creation:

\begin{enumerate}
    \item \textbf{AI Generative Architect.} An intent-aware prompt generation system that supports persona switching across multiple large language model (LLM) providers. The architect interprets the user's pedagogical intent and generates a structured prompt with embedded \texttt{[[variable\_name]]} parameter syntax---using double brackets to avoid collision with LaTeX's single-bracket commands.

    \item \textbf{Scholarly Inject Toolkit.} A paste-and-parse interface where raw AI output is injected into PEDAL, triggering automated extraction of 24+ metadata fields. The toolkit parses the output to identify Bloom's level, prompting strategy, audience level, subject domain, and evaluation metrics, populating the SS2 metadata envelope automatically.

    \item \textbf{Artifact Ingest.} A ``PDF-to-prompt bootstrap'' workflow designed to process existing educational materials (e.g., textbook chapters, lab protocols). The interface features a PDF export button that copies a tailored meta-prompt, dynamically generated based on the current SS2 metadata parameters configured via the UI (such as switching Bloom's Level from \textit{Analyze} to \textit{Apply} using dropdowns). The researcher pastes this tailored meta-prompt alongside the source PDF document into an external LLM. The resulting generated prompt is ingested back into PEDAL as a new, immutable Git-style version (e.g., v2). This variant is then queued for Laboratory evaluation, where the \texttt{LAB PREFERRED} statistical engine seamlessly compares prompt effectiveness across both vertical version histories and horizontal branches.
\end{enumerate}

\begin{figure}[htbp]
    \centering
    \includegraphics[width=0.9\textwidth, height=0.8\textheight, keepaspectratio]{pedal-2-editor-view}
    \caption{PEDAL Orchestration Layer: The Editor View showing the 32-field metadata entry system, the versioning branch selector, and the prompt narrative construction area.}
    \label{fig:editor_view}
\end{figure}

The output of Layer~1 is a \textit{Master Artifact}: a prompt record with an initial version (v1), a populated metadata envelope, and a status of \texttt{draft}.

\subsection{Layer 2: Laboratory}
\label{sec:layer2}

The Laboratory layer (\texttt{dossier/\allowbreak prompt\_edit.php}, \texttt{dossier/\allowbreak prompt\_test.php}) supports iterative refinement through three capabilities:

\begin{enumerate}
    \item \textbf{Git-Style Version Control.} Every prompt maintains a tree structure with named branches. New versions are created as commits on a branch, with each version receiving an immutable snapshot. The system tracks parent--child relationships, enabling dependency graph visualization and merge operations between branches. Historical versions are read-only once committed, ensuring audit trail integrity.

    \item \textbf{LLM-as-a-Judge Evaluation.} A standardized testing environment utilizing a pure AI evaluation model, with no human scoring involved. After a research artifact is generated, the researcher copies a custom evaluation prompt dynamically constructed by PEDAL and pastes it into an external LLM alongside the artifact's output. The LLM acts as the sole judge, evaluating the output across five dimensions (Bloom's Alignment, Socratic Integrity, Subject Precision, Persona Consistency, and Actionability) and returning the results as structured JSON. The researcher injects this JSON back into PEDAL, which parses the data, calculates an aggregate quality score out of 5.0, and flags the ``Lab Preferred'' variant.

    \item \textbf{Research Lab.} A single-architecture testing view (\texttt{prompt\_test.php}) where prompt variants are executed with \texttt{[[variable]]} parameter injection. Because these \texttt{[[variable]]} parameters often encompass complex datasets---where extreme detail yields superior AI scaffolding for the learner---researchers currently use an external prompt to generate meaningful inputs and update these fields manually. A future software update is scheduled to embed this variable-generation pipeline directly into the PEDAL framework. A highly detailed example of an \texttt{[[Execution\_Protocol]]} variable is as follows:

\begin{quote}
\small
\textbf{[[Execution\_Protocol]]} \\
Phase 1: Real-World Hook. Present a scenario: ``A forensic tech finds an unlabeled bottle of vinegar. Titration shows it's 0.8M Acetic Acid. If the legal standard is 5\% mass/volume, is this batch legal?'' Phase 2: Variable Audit. Ask: ``Before we calculate, what is the `target unit' for the legal standard vs.\ our lab result? What bridge (conversion factor) connects them?'' Phase 3: The `Mole' Bridge. Scaffolding: ``We have 0.8 moles in 1 Liter. What is the mass of 0.8 moles of $CH_3COOH$?'' (Provide Atomic Weights: $C=12.01, H=1.008, O=16.00$). Phase 4: Volume/Density Check. Ask: ``If the density is approximately $1.00 \text{ g/mL}$, how does that change our perception of the 1 Liter of solution's total mass?'' Phase 5: Evaluation. Ask the student to compare the calculated mass \% to the 5\% legal limit. \citep{kahveci2026-cp}
\end{quote}

The Research Lab relies on a manual ``Copy \& Log'' workflow: the user copies the fully resolved prompt, executes it in an external LLM interface, and logs the structured result back into PEDAL. This deliberate human-in-the-loop design ensures that PEDAL never makes unsupervised API calls to LLM providers, preserving editorial control and model-agnosticism.
\end{enumerate}

\begin{figure}[htbp]
    \centering
    \includegraphics[width=0.9\textwidth, height=0.8\textheight, keepaspectratio]{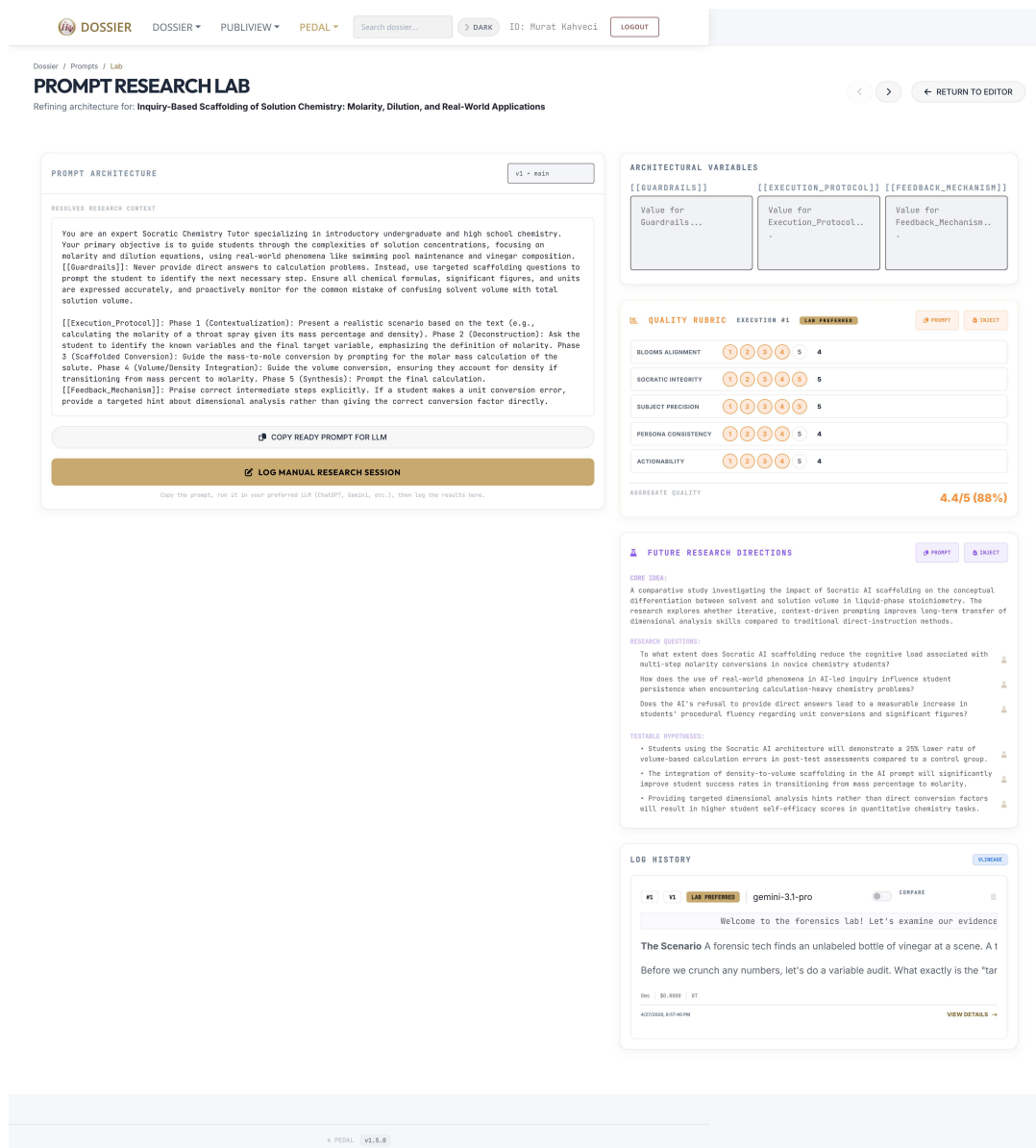}
    \caption{PEDAL Laboratory Layer: The Research Lab interface showing a prompt execution session, multi-dimensional rubric scoring, and future research direction logging.}
    \label{fig:lab_view}
\end{figure}

The output of Layer~2 is a set of evaluated versions with quality scores, rubric breakdowns, and statistical comparisons. The highest-scoring variant is automatically identified as \textbf{LAB PREFERRED} (see \S\ref{sec:stats}).

\subsection{Layer 3: Public Archive}
\label{sec:layer3}

The Public Archive layer (\texttt{pub/\allowbreak prompt.php}, \texttt{pub/\allowbreak prompts.php}) transforms laboratory-validated prompts into discoverable scholarly artifacts:

\begin{enumerate}
    \item \textbf{PubliView.} A scholarly public interface that displays the artifact with full metadata, version navigation, statistical summaries, and citation information. The system auto-loads the highest-scoring version by default (``statistics-first selection''), ensuring that visitors see the most rigorously validated variant.

    \item \textbf{Citation Engine.} One-click citation export in formats optimized for scholarly use, including APA and BibTeX. Each citation carries the immutable per-version DOI, ensuring that prompt references remain persistent and verifiable.

    \item \textbf{Multi-Format Data Export.} Machine-readable download of the complete SS2 metadata envelope and pedagogical specifications. The system supports three primary export formats: (a)~\textbf{JSON} for programmatic consumption and archival integrity; (b)~\textbf{CSV} for large-scale systematic reviews and spreadsheet-based analysis; and (c)~\textbf{\LaTeX} snippets for direct inclusion in technical manuscripts.

    \item \textbf{SEO and Discoverability.} The archive implements three discoverability strategies: (a)~JSON-LD \texttt{Dataset} schema for Google Dataset Search indexing; (b)~HighWire Press meta headers for Google Scholar visibility; and (c)~a dynamic XML sitemap using the \texttt{pub\_slugs} registry as the single source of truth, with per-artifact last-modified timestamps to trigger crawler re-indexing.
\end{enumerate}

\subsection{Technology Stack Summary}

Table~\ref{tab:stack} summarizes the platform's technology stack.

\begin{table}[htbp]
\centering
\caption{PEDAL technology stack (v1.5.0)}
\label{tab:stack}
\footnotesize
\begin{tabularx}{\textwidth}{lX}
\toprule
\textbf{Layer} & \textbf{Technology} \\
\midrule
Backend & PHP 8.x, MySQL/MariaDB 11.8 \\
Frontend & Vanilla PHP/JS, Bootstrap 5 \\
Design system & True Dark Industrial (\texttt{\#0D0D0D} bg, \texttt{\#141414} cards, \texttt{\#1F1F1F} borders) \\
Typography & Outfit (headings), Inter (body), JetBrains Mono (code) \\
DOI minting & Zenodo REST API (\texttt{ZenodoService} class) \\
SEO & JSON-LD Dataset schema, HighWire Press meta, \texttt{pub\_slugs} sitemap \\
Rendering & KaTeX (LaTeX), Mermaid (diagrams), Markdown \\
Version & Dossier v3.5.0, PEDAL v1.5.0 \\
\bottomrule
\end{tabularx}
\end{table}
% =============================================================================
% §4  SCHOLARLY SYNC 2 & RESEARCH CLASSIFICATION
% =============================================================================
\section{Scholarly Sync 2 and Research Classification}
\label{sec:ss2}

\subsection{The SS2 Metadata Framework}

Scholarly Sync~2 (SS2) is PEDAL's coordination protocol for ensuring that every prompt artifact carries a comprehensive, machine-readable metadata envelope. The framework operates on three core principles: (1)~\textit{automated propagation}---status changes cascade to public DOI records automatically; (2)~\textit{audit trail integrity}---every sync event is logged with a timestamp and author UUID; and (3)~\textit{machine-readable exports}---auto-generated JSON research snapshots, \texttt{.csv} datasets, and \LaTeX{} source code accompany every public artifact.

Each artifact's SS2 metadata JSON block encodes the key sections detailed in Table~\ref{tab:ss2_metadata}.

\begin{table}[htbp]
\centering
\caption{Core SS2 metadata JSON payload structure}
\label{tab:ss2_metadata}
\footnotesize
\begin{tabularx}{\textwidth}{lX}
\toprule
\textbf{JSON Key} & \textbf{Description} \\
\midrule
\texttt{artifact\_metadata\_version} & Schema version control for the metadata envelope. \\
\texttt{pedagogical\_intent} & Primary educational purpose of the prompt. \\
\texttt{cognitive\_taxonomy} & Bloom's Revised Taxonomy and Webb's DOK mappings. \\
\texttt{pedagogical\_frameworks} & 5E Instructional Model and SAMR integration levels. \\
\texttt{standards\_alignment} & NGSS standards, including SEP, CCC, and DCI codes. \\
\texttt{misconception\_targets} & Anticipated student misconceptions addressed. \\
\texttt{prompt\_specifications} & Target model, system prompt, and execution parameters. \\
\texttt{evaluation\_metrics} & Five-dimension rubric scoring data. \\
\texttt{modality\_specifications} & Modality constraints (e.g., text, image, code). \\
\texttt{evaluation\_status} & Current validation state (e.g., LAB PREFERRED). \\
\texttt{licensing\_and\_attribution} & CC-BY-4.0 licensing and author attribution. \\
\texttt{archival\_metadata} & Zenodo DOI, timestamps, and platform citation data. \\
\bottomrule
\end{tabularx}
\end{table}

Platform citation metadata is centrally maintained in \texttt{data/\allowbreak pedal\_citation.json}.

\subsection{Cognitive Taxonomy Integration}

PEDAL classifies every artifact along two cognitive dimensions:

\begin{enumerate}
    \item \textbf{Bloom's Revised Taxonomy} \citep{anderson2001}: Six levels from \textit{Remember} through \textit{Create}, based on the seminal work of Bloom et al. \citep{bloom1956}. Each prompt is tagged with the highest cognitive level it is designed to elicit from the learner.
    \item \textbf{Webb's Depth of Knowledge (DOK)} \citep{webb1997}: Four levels from \textit{Recall \& Reproduction} (Level~1) through \textit{Extended Thinking} (Level~4). DOK captures the complexity of thinking required rather than the verb taxonomy of Bloom's, providing a complementary classification axis.
\end{enumerate}

\subsection{Pedagogical Framework Mapping}

Two instructional design frameworks are encoded per artifact:

\begin{enumerate}
    \item \textbf{SAMR Model} \citep{puentedura2006}: Classifies the technology integration level as Substitution, Augmentation, Modification, or Redefinition. This enables researchers to filter the archive by the degree of pedagogical transformation a prompt facilitates.
    \item \textbf{5E Instructional Model} \citep{bybee2006}: Maps each prompt to one or more phases of the BSCS inquiry cycle: Engage, Explore, Explain, Elaborate, or Evaluate. This classification supports curriculum-aligned prompt sequencing.
\end{enumerate}

\subsection{NGSS Standards Alignment}

For science education prompts, PEDAL maintains a dedicated \texttt{prompt\_\allowbreak standards\_\allowbreak alignment} table linking artifacts to NGSS standards \citep{ngss2013}. Each alignment record encodes: the standard code, description, Science \& Engineering Practice (SEP), Crosscutting Concept (CCC), and Disciplinary Core Idea (DCI) code with description. A master registry (\texttt{ngss\_master}) provides the authoritative standards catalog against which alignments are validated.

\subsection{Dual-Track Classification}

PEDAL implements a dual-track classification system that distinguishes between prompt-level and version-level metadata:

\begin{itemize}
    \item \textbf{Prompt-level classification} (stored on the \texttt{prompts} table): Captures invariant properties of the artifact---subject domain, audience level, pedagogical intent, textbook alignment, and NGSS standards. These properties persist across all versions.
    \item \textbf{Version-level classification} (stored on \texttt{prompt\_versions}): Captures properties that may change between iterations---Bloom's level, prompting strategy, default model, and system prompt. This enables researchers to track how cognitive level or prompting approach evolves across refinement cycles.
\end{itemize}

Table~\ref{tab:classification} summarizes the complete classification grid.

\begin{table}[htbp]
\centering
\caption{Research classification grid with enumerated values}
\label{tab:classification}
\footnotesize
\begin{tabularx}{\textwidth}{llX}
\toprule
\textbf{Dimension} & \textbf{Track} & \textbf{Values} \\
\midrule
Bloom's level & Version & remember, understand, apply, analyze, evaluate, create \\
DOK level & Prompt & 1\_recall, 2\_skill\_concept, 3\_strategic\_thinking, 4\_extended\_thinking \\
Pedagogical intent & Prompt & lesson\_planning, assessment\_generation, feedback\_scaffolding, inquiry\_facilitation, content\_explanation, differentiation, metacognition, lab\_design, discussion\_facilitation, rubric\_generation, misconception\_diagnosis, other \\
Audience level & Prompt & k5, middle\_school, high\_school, undergraduate, graduate, professional, general \\
Prompting strategy & Version & zero\_shot, few\_shot, chain\_of\_thought, role\_based, constrained, scaffolded, socratic, tree\_of\_thought, iterative, rag\_augmented, multi\_turn, other \\
SAMR level & Prompt & substitution, augmentation, modification, redefinition \\
5E phase & Prompt & engage, explore, explain, elaborate, evaluate \\
Output format & Prompt & free\_text, rubric, quiz, lesson\_plan, lab\_protocol, discussion\_questions, code, data\_table, cer\_format, worked\_example, case\_study, other \\
\bottomrule
\end{tabularx}
\end{table}
% =============================================================================
% §5  STATISTICAL ENGINE & ZENODO DOI INTEGRATION
% =============================================================================
\section{Statistical Engine and DOI Integration}
\label{sec:stats}

\subsection{Mann--Whitney \texorpdfstring{$U$}{U} Test Implementation}

PEDAL employs the Mann--Whitney $U$ test \citep{mann1947}---a non-parametric rank-sum test---as its primary statistical engine for comparing prompt variant performance. This choice is deliberate: prompt evaluation scores are ordinal (1--5 Likert scale), sample sizes per variant are typically small ($n < 30$), and the distribution of quality scores is rarely normal. The implementation (\texttt{ResearchStats::mannWhitneyU} in \texttt{includes/\allowbreak ResearchStats.php}) includes tie correction via average rank assignment, a variance correction factor, and $Z$-score computation using normal approximation with the Abramowitz \& Stegun cumulative distribution function.

The significance threshold is set at $p < 0.1$ (relaxed from the conventional 0.05) to accommodate the small sample sizes inherent in iterative prompt engineering. This methodological choice is documented explicitly in the platform and in all statistical outputs.

\subsection{LAB PREFERRED Criteria}

The LAB PREFERRED designation identifies the highest-performing variant in each prompt's version tree. A version earns this badge when it satisfies two criteria simultaneously:

\begin{enumerate}
    \item \textbf{Highest average quality score} across all versions in the lineage (computed from the five rubric dimensions).
    \item \textbf{Statistical significance} ($p < 0.1$ via Mann--Whitney $U$) against the baseline version (v1 or the earliest version with empirical evaluation data).
\end{enumerate}

The LAB PREFERRED badge is displayed in the Laboratory view, propagated to the PubliView, and included in the artifact's citation metadata. When a new version surpasses the current LAB PREFERRED variant, the badge transfers automatically.

\subsection{Rubric Scoring System}

Each prompt execution is evaluated across five default rubric dimensions, stored in the \texttt{prompt\_quality\_rubric} table:

\begin{enumerate}
    \item \textbf{Bloom's Alignment} (\texttt{blooms\_alignment}): Does the prompt elicit the targeted cognitive level?
    \item \textbf{Socratic Integrity} (\texttt{socratic\_integrity}): Does the prompt scaffold learning without providing direct answers?
    \item \textbf{Subject Precision} (\texttt{subject\_precision}): Is the domain-specific content accurate?
    \item \textbf{Persona Consistency} (\texttt{persona\_consistency}): Does the prompt maintain its instructional tone and pedagogical intent?
    \item \textbf{Actionability} (\texttt{actionability}): Can a practitioner implement the prompt output in a real classroom setting?
\end{enumerate}

Each dimension is scored on a 1--5 scale (Insufficient to Exceptional). The aggregate quality score is the unweighted mean across all five dimensions. Scores, rater identity, rater role, and free-text notes are stored per execution, enabling inter-rater reliability analysis.

\subsection{Zenodo Integration Architecture}

PEDAL's DOI minting is implemented through the \texttt{ZenodoService} class (\texttt{includes/\allowbreak zenodo\_service.\allowbreak php}), which interfaces 
with the Zenodo REST API \citep{zenodo2013}. The deposit workflow proceeds in five steps:

\begin{enumerate}
    \item \textbf{Create deposition} (or a new version of an existing concept DOI).
    \item \textbf{Upload prompt text} as a \texttt{.txt} file.
    \item \textbf{Upload JSON metadata sidecar} containing the full SS2 envelope.
    \item \textbf{Set Zenodo metadata} (creators, description, license, keywords).
    \item \textbf{Publish} and receive the version-specific DOI.
\end{enumerate}

Three DOI fields are stored per artifact: \texttt{doi} (version-specific), \texttt{zenodo\_id} (deposition ID for updates), and \texttt{zenodo\_concept\_doi} (version-agnostic concept DOI that always resolves to the latest version).

\subsection{Multi-Author DOI Support}

The Zenodo deposit system queries the \texttt{prompt\_authors} table for ordered author lists. For collaborative prompts, the first author (collaborator) appears as Creator~1, the PI as Creator~2, with ORCID and institutional affiliations included when available. Only prompts with \texttt{approval\_status = 'approved'} can be deposited to Zenodo, ensuring that all DOI-minted artifacts have passed the PI quality gate.
% =============================================================================
% §6  COLLABORATIVE ROLE-BASED ACCESS CONTROL
% =============================================================================
\section{Collaborative Access Control}
\label{sec:rbac}

\subsection{Role Architecture}

PEDAL implements a two-role access control system designed for research group hierarchies:

\begin{itemize}
    \item \textbf{Principal Investigator (PI):} Full platform access including all CRUD operations, Zenodo DOI deposits, system settings, collaborator account management, and approval authority. The PI gate ensures that no artifact reaches public visibility or DOI status without senior researcher endorsement.
    \item \textbf{Collaborator:} Scoped exclusively to the PEDAL Archive. Collaborators can create and edit their own prompts, execute evaluations, and submit prompts for PI review. They cannot access other platform modules, manage settings, or directly publish artifacts.
\end{itemize}

Role enforcement is implemented through helper functions in \texttt{includes/\allowbreak bootstrap.php}: \texttt{is\_pi()}, \texttt{is\_collaborator()}, and \texttt{require\_pi()} (which blocks non-PI access with a redirect or HTTP 403).

\subsection{Approval Workflow}

The approval pipeline ensures quality control over collaborative artifacts within the {PEDAL} framework:

\begin{enumerate}[leftmargin=*, labelsep=1em]
    \item \textbf{Creation:} Collaborator creates a prompt: 
          \hfill \texttt{\small approval\_status = 'not\_\allowbreak submitted'}
    \item \textbf{Submission:} Collaborator submits for review: 
          \hfill \texttt{\small approval\_status = 'pending\_\allowbreak review'}
    \item \textbf{PI Review:} \texorpdfstring{{PI}}{PI} evaluates the artifact and either approves (\texttt{'approved'}) or rejects with notes (\texttt{'rejected'}). Revision requests are also supported.
    \item \textbf{Publication Gate:} Only prompts with \texttt{approval\_status = 'approved'} can be set to \texttt{production} status or deposited to {Zenodo} for {DOI} minting.
\end{enumerate}

All approval actions are logged in the \texttt{prompt\_approvals} table with timestamps, reviewer identity, and PI notes for audit trail purposes.

\subsection{Multi-Author Attribution}

The \texttt{prompt\_authors} table supports ordered multi-author attribution with explicit roles: \texttt{first\_author}, \texttt{pi}, and \texttt{co\_author}. Author order is preserved in all citation exports and Zenodo deposits. Each author record includes \texttt{user\_id}, \texttt{author\_order}, and \texttt{role}, enabling accurate academic credit allocation.
% =============================================================================
% §7  IMPLEMENTATION EXEMPLAR 1: CHEMISTRY EDUCATION
% =============================================================================
\section{Implementation Exemplar 1: Chemistry Education Domain}
\label{sec:chemistry_exemplar}

While the PEDAL architecture supports broad applications, the ``Chemistry Education Domain Exemplar'' serves as a foundational application of the Scholarly Sync 2 (SS2) framework within Chemistry Education---specifically focusing on Solution Chemistry, Molarity, and Dilution.

\subsection{Technical Mapping and SS2 Injection}

This exemplar utilizes the platform's Scholarly Inject toolkit to automatically parse over 24 distinct metadata fields from raw chemistry problem outputs. When the prompt is evaluated against an LLM to generate inquiry-based scaffolding scenarios, the resulting output is systematically processed to populate the multidimensional SS2 metadata envelope. 

Crucially, the system automatically classifies the artifact's pedagogical parameters:
\begin{itemize}
    \item \textbf{Cognitive Domain:} Mapped to Bloom's Taxonomy at the \textit{Create} level.
    \item \textbf{Rigor Level:} Classified as Depth of Knowledge (DOK) Level 4 (Extended Thinking).
    \item \textbf{Standards Alignment:} Formally anchored to NGSS Performance Expectation \textit{HS-PS1-7} (Use mathematical representations to support the claim that atoms, and therefore mass, are conserved during a chemical reaction).
\end{itemize}

\subsection{Research Scaffolding}

By structuring the artifact with formal metadata, the prompt transcends its origin as a ``disposable instruction'' and becomes a ``citable research instrument'' suitable for rigorous academic inquiry, such as a Master's Thesis. While PEDAL is foundational for educational scaffolding, it is equally designed as a robust infrastructure for STEM researchers. This is best evidenced by the platform's research generation capabilities: the \texttt{future\_directions} segment actively drives scholarly output by proposing novel research ideas, formulating three specific research questions on the target topic, and positing three testable hypotheses for empirical validation.

The system scaffolds this research intent through dedicated metadata fields, representing the core data tracked in the platform's research log (Table~\ref{tab:research_scaffolding}). A full visual rendering of this chemistry exemplar within the platform's PubliView interface is provided in Figure~\ref{chemistry-examplary} (Appendix~\ref{app:gallery}). The complete prompt artifact, including its system parameters and version history, is persistently archived and formally citable via its DOI \citep{kahveci2026-cp}.

\begin{table}[htbp]
\centering
\caption{Research scaffolding for the Chemistry Education Domain Exemplar}
\label{tab:research_scaffolding}
\begin{tabularx}{\textwidth}{lX}
\toprule
\textbf{SS2 Field} & \textbf{Chemistry Exemplar Content} \\
\midrule
\texttt{research\_questions} & How do variations in prompt structure impact the ability of AI to model inquiry-based learning conversations about solution chemistry? \\
\texttt{misconception\_targets} & Confusing Molarity (M) with Molality (m); incorrectly applying the dilution equation ($M_1V_1 = M_2V_2$); misinterpreting moles of solute vs.\ volume of solution. \\
\texttt{future\_directions} & Evaluate the prompt's efficacy across diverse undergraduate cohorts to assess its impact on mitigating persistent molarity misconceptions. \\
\bottomrule
\end{tabularx}
\end{table}

% =============================================================================
% §8  IMPLEMENTATION EXEMPLAR 2: NExAIE PEER REVIEW
% =============================================================================
\section{Implementation Exemplar 2: The NExAIE Extension}
\label{sec:nexaie}

To demonstrate the extensibility and production-readiness of the PEDAL infrastructure, we present NExAIE (Nexus AI \& Education)---a specialized module that applies the Laboratory-to-Archive pipeline to the challenge of AI-augmented peer review. NExAIE is not a standalone application but a conceptual and technical extension of the PEDAL core, inheriting its versioning, metadata, and security protocols. This section details how NExAIE utilizes the platform's architectural layers; a comprehensive methodology and empirical evaluation of the NExAIE review framework are provided in a companion paper \citep{kahveci2026nexaie}.

\subsection{Design Principles}

Five non-negotiable architectural principles govern the NExAIE design:

\begin{enumerate}[label=\textbf{P\arabic*.}]
    \item \textbf{OJS-First Management.} The Open Journal Systems (OJS) installation \citep{willinsky2005} retains exclusive authority over manuscript submission, author--editor communication, and formal Accept/Reject decisions. PEDAL never receives, stores, or processes submitted manuscripts directly.

    \item \textbf{Copy \& Log Orchestration.} PEDAL does not make direct API calls to any LLM provider. The Editor-in-Chief (EiC) manually copies a version-controlled review prompt from PEDAL, executes it against the manuscript using a professional-grade LLM, and injects the structured JSON response back into PEDAL. This human-in-the-loop design ensures editorial judgment at every step and maintains model-agnosticism.

    \item \textbf{Confidentiality by Default.} All identifying metadata---manuscript title, author names, and institutional affiliations---is encrypted at rest using AES-256-CBC \citep{nist2001}. The public-facing view displays only a permanent pseudocode identifier (e.g., \texttt{NXR-2026-0042}).

    \item \textbf{Dual-View Rendering.} Every Paper Review record has exactly two renderings: the \textit{Editor View} (full detail, authenticated) and the \textit{Public View} (pseudocoded, dimension scores only, accessible to anyone).

    \item \textbf{Citable Provenance.} Every review record and the protocol prompts used to generate it receives a DOI via Zenodo, ensuring that published articles can cite their exact review methodology.
\end{enumerate}

\subsection{System Boundary: OJS--PEDAL Integration}

Table~\ref{tab:boundary} specifies the strict separation of concerns.

\begin{table}[htbp]
\centering
\caption{System boundary between OJS and PEDAL}
\label{tab:boundary}
\begin{tabularx}{\textwidth}{lXX}
\toprule
\textbf{Function} & \textbf{OJS} & \textbf{PEDAL Archive} \\
\midrule
Manuscript submission & \checkmark & --- \\
Author--Editor communication & \checkmark & --- \\
Final Accept/Reject decision & \checkmark & --- \\
Published article hosting & \checkmark & --- \\
Review protocol prompts (v/c, DOI) & --- & \checkmark \\
Paper review records (pseudocoded) & --- & \checkmark \\
AI review execution logging & --- & \checkmark \\
DOI minting for review artifacts & --- & \checkmark \\
\bottomrule
\end{tabularx}
\end{table}

The bidirectional link between systems is established through the \textit{Article AI Artifact} citation: the OJS article includes an ``AI-Human Review Disclosure'' block referencing the PEDAL protocol slug and version, while the PEDAL record stores the published article's DOI.

\subsection{The 42-Prompt Evaluation Matrix}

The NExAIE methodology decomposes manuscript evaluation into six orthogonal dimensions, each assessed by a dedicated, publication-type-specific prompt. Rather than relying on holistic evaluation, the framework produces six independent scores with structured evidence. The rubrics were calibrated through systematic synthesis of reviewer guidelines from four top-tier journals:

\begin{itemize}
    \item \textit{Journal of Research in Science Teaching} (JRST): ``Extend current knowledge or provide new perspectives''
    \item \textit{Computers \& Education}: ``Significant contribution to current knowledge''; measurable learning outcomes mandatory
    \item \textit{British Journal of Educational Technology} (BJET): ``Novel perspective, new theoretical insights''
    \item \textit{International Journal of Science Education} (IJSE): ``Contribute new knowledge to the field''
\end{itemize}

\subsection{Six Core Dimensions}

All manuscripts are evaluated on a 1--5 scale across six dimensions:

\begin{enumerate}
    \item \textbf{Academic Merit \& Originality} (\texttt{academic\_merit}): Novelty, theoretical framing, significance. Key differentiator: ``adding another brick'' (3) vs.\ ``showing the wall could be built differently'' (4).
    \item \textbf{Methodological \& Statistical Rigor} (\texttt{methodological\_rigor}): Study design, reproducibility, validity. Key differentiator: proactive validity threats with effect sizes (4) vs.\ standard methods (3).
    \item \textbf{Data Integrity \& Interpretation} (\texttt{data\_integrity}): Evidence-to-conclusion alignment, limitation honesty. Key differentiator: dismissing alternatives with evidence (4) vs.\ reasonable conclusions without exploration (3).
    \item \textbf{Pedagogical Alignment} (\texttt{pedagogical\_alignment}): Learning theory grounding, equity considerations. Key differentiator: theory \textit{operationalized} (4) vs.\ theory \textit{cited} (3).
    \item \textbf{Structural Quality} (\texttt{structural\_quality}): Logical flow, writing clarity. Key differentiator: effortless reading (4) vs.\ reader effort required (3).
    \item \textbf{Citation Compliance} (\texttt{citation\_compliance}): Coverage, currency ($\geq$60\% from last 5 years), critical engagement. Key differentiator: \textit{engaging} with sources (4) vs.\ \textit{listing} them (3).
\end{enumerate}

The universal scoring rubric defines five levels: Exemplary~(5), Strong~(4), Adequate~(3), Weak~(2), and Insufficient~(1). Score~5 is calibrated as genuinely rare ($<$10\% of reviews), while Score~3 represents competent, publishable-with-revisions work.

\subsection{Publication Type Adaptation}

The universal rubrics are adapted for seven publication types. Table~\ref{tab:matrix} summarizes the key reframings.

\begin{table}[htbp]
\centering
\caption{Publication type adaptation matrix}
\label{tab:matrix}
\footnotesize
\begin{tabularx}{\textwidth}{lXXX}
\toprule
\textbf{Type} & \textbf{Methodology Reframing} & \textbf{Primary Emphasis} & \textbf{Special Rule} \\
\midrule
Research & Standard (IMRaD, effect sizes) & All dimensions equal & No descriptive reports \\
Review & Search strategy, PRISMA & Citations: exhaustive & Not a literature dump \\
STEM Ed & STEM instruments, pre/post & Pedagogical: NGSS & Must address equity \\
Tech Report & Technical validation, WCAG & Methodology: benchmarks & Not a product brochure \\
Communication & Single method, robust & Merit: urgency & Brevity is a virtue \\
Lab Experiment & Pre/post assessment & Pedagogical: materials & Must be implementable \\
Commentary & Argumentative rigor & Merit \& Structure & Opinion $\neq$ evidence \\
\bottomrule
\end{tabularx}
\end{table}

\subsection{Prompt Architecture and JSON Schema}

Each of the 42 dimension-specific prompts follows a standardized internal structure: (1)~system identity and journal context; (2)~single-dimension focus instruction; (3)~anti-inflation calibration directives; (4)~publication-type-specific 5-level rubric; (5)~red flag triggers; (6)~JSON output schema; and (7)~Socratic public note instructions.

Every dimension evaluation must return a valid JSON object containing: \texttt{dimension}, \texttt{score} (integer 1--5), \texttt{confidence} (float 0.0--1.0), \texttt{summary} (3--5 sentences), \texttt{strengths} and \texttt{weaknesses} (arrays, minimum 20 characters per item), \texttt{specific\_evidence} (array of claim/quote/section triples), \texttt{recommendations}, \texttt{red\_flags}, and \texttt{public\_note}.

\subsection{Synthesis Engine}

After all six dimension JSONs are injected, a universal Synthesis prompt consumes the aggregate data to produce a holistic assessment. The synthesis output is structured into the following metadata fields:

\begin{description}[nosep, leftmargin=1em]
    \item[\texttt{overall\_quality\_note}] Holistic pedagogical assessment.
    \item[\texttt{scope\_fit\_note}] Alignment with journal/repository scope.
    \item[\texttt{suggested\_verdict}] Automated recommendation (\textit{Accept/Reject}).
    \item[\texttt{key\_strengths/concerns}] Bulleted highlights of the artifact.
    \item[\texttt{revision\_guidance}] Actionable Socratic feedback for authors.
\end{description}

Verdict calibration follows explicit thresholds: aggregate $\geq 4.5$ with no dimension below 3 suggests ``accept''; $\geq$2 dimensions scoring $\leq$2 suggests ``reject.'' Override conditions ensure that a score of 1 in either Academic Merit or Methodological Rigor triggers a ``reject'' advisory regardless of aggregate performance.
\subsection{Code and Security Integration}
\label{sec:nexaie_security}

Because NExAIE represents a complete, auditable peer review record, author anonymity and data integrity are paramount. Confidential metadata for a reviewed manuscript---such as the principal investigator's identity and institutional affiliation---is processed using the platform's AES-256-CBC encryption logic before being stored in \texttt{VARBINARY} columns, as defined in Section~\ref{sec:data}. 

When a review protocol prompt is executed in the Research Lab, dynamic variables are safely substituted without exposing sensitive state to external observers. For example, the system parses syntax such as \texttt{[[manuscript\_title]]} and \texttt{[[abstract]]} within the prompt narrative. It executes the injection, evaluates the output, and securely encrypts the resulting \texttt{author\_notes} and \texttt{pi\_feedback} fields. This dual-layer architecture ensures that the peer review protocol can be openly cited and replicated via its DOI, while the underlying researcher identities remain fully protected from unauthorized access.

% =============================================================================
% §8  RADICAL TRANSPARENCY & DUAL-VIEW ARCHITECTURE
% =============================================================================
\section{Radical Transparency and Dual-View Architecture}
\label{sec:transparency}

\subsection{The Radical Transparency Paradigm}

NExAIE introduces \textit{Radical Transparency}: the complete public disclosure of the evaluation methodology, including the exact prompts, rubrics, and scoring criteria used to assess manuscripts. All 42 dimension-specific prompts and the synthesis prompt are published as version-controlled PEDAL artifacts, each minted with a DOI via Zenodo. A dedicated public-facing portal allows anyone---including prospective authors---to browse, copy, and use these prompts for pre-submission self-review.

This design serves three purposes: (1)~\textit{author empowerment}---authors can identify and address weaknesses before submission; (2)~\textit{evaluation reproducibility}---any researcher can independently verify the rigor of NExAIE's review process; and (3)~\textit{scholarly contribution}---the rubrics themselves become citable research instruments that other journals may adopt or adapt.

\subsection{Socratic Public Notes}

Each dimension evaluation produces a \texttt{public\_note}: a 3--5 sentence passage written in the voice of a Socratic graduate mentor. These notes are carefully constrained: they must not reference the paper's title, authors, or specific findings; they address the \textit{dimension} broadly rather than critiquing the specific manuscript; and they use forward-looking framing (``One might consider how\ldots'') rather than criticism. These notes appear on the Public View alongside dimension scores, providing intellectual context without compromising confidential review details.

\subsection{Dual-View Data Visibility}

Table~\ref{tab:dualview} specifies which data elements are visible in each rendering mode.

\begin{table}[htbp]
\centering
\caption{Data visibility: Editor View vs.\ Public View}
\label{tab:dualview}
\footnotesize
\begin{tabularx}{\textwidth}{Xcc}
\toprule
\textbf{Data Element} & \textbf{Editor View} & \textbf{Public View} \\
\midrule
Real paper title & \checkmark & --- (pseudocode) \\
Author names / Institution & \checkmark & --- \\
All 6 dimension scores & \checkmark & \checkmark (scores only) \\
Dimension narratives & \checkmark & --- \\
Strengths / Weaknesses & \checkmark & --- \\
Specific evidence (quotes) & \checkmark & --- \\
Recommendations & \checkmark & --- \\
Red flags & \checkmark & --- \\
Socratic public notes & \checkmark & \checkmark \\
Aggregate score \& Verdict & \checkmark & \checkmark \\
Synthesis quality \& scope notes & \checkmark & \checkmark \\
Protocol reference (PEDAL slug) & \checkmark & \checkmark \\
Model used \& review date & \checkmark & \checkmark \\
Review DOI \& citation export & \checkmark & \checkmark \\
Article DOI (if accepted) & \checkmark & \checkmark \\
\bottomrule
\end{tabularx}
\end{table}

% =============================================================================
% §9  DATA MODEL & SECURITY
% =============================================================================
\section{Data Model and Security}
\label{sec:data}

\subsection{Database Schema Extensions}

The NExAIE extension adds 13 columns to the existing PEDAL \texttt{prompts} table 
rather than creating separate review tables, preserving compatibility with 
PEDAL's existing version control, DOI minting, and rendering infrastructure. 
Key additions include: \texttt{is\_\allowbreak nexaie\_\allowbreak review} 
(boolean flag), \texttt{review\_\allowbreak pseudocode} (VARCHAR(20), unique), 
\texttt{review\_\allowbreak publication\_\allowbreak type} (ENUM of seven types), 
\texttt{review\_data} (LONGTEXT JSON), \texttt{review\_\allowbreak aggregate\_\allowbreak score} 
(DECIMAL(3,2)), \texttt{review\_verdict} (ENUM), \texttt{review\_\allowbreak synthesis} 
(LONGTEXT JSON), \texttt{editor\_notes\_public}, three \texttt{VARBINARY} columns 
for encrypted confidential metadata, \texttt{article\_doi}, and \texttt{share\_token}. 
A separate \texttt{nexaie\_\allowbreak review\_\allowbreak sequence} table maintains auto-incrementing 
counters per year for pseudocode uniqueness.

\subsection{Pseudocode System}

Each Paper Review record receives a permanent pseudocode in the format \texttt{NXR-YYYY-NNNN}, where \texttt{NXR} denotes ``NExAIE Review,'' \texttt{YYYY} is the submission year, and \texttt{NNNN} is a zero-padded sequential number. This pseudocode serves as the public-facing identifier, URL slug, and citable reference. Once assigned, a pseudocode is never changed or reused.

\subsection{Encryption at Rest}

Confidential metadata is encrypted using AES-256-CBC \citep{nist2001} with a server-side key stored outside the web root. The encryption workflow: (1)~the EiC provides real title, author names, and institution at record creation; (2)~values are encrypted using \texttt{openssl\_encrypt()} with a random initialization vector (IV); (3)~the IV is prepended to the ciphertext and stored in \texttt{VARBINARY} columns; (4)~decryption occurs only during authenticated Editor View rendering; (5)~the Public View rendering path never calls the decryption function.

\subsection{Share Token Access}

Section Editors receive read-only access via a URL containing a 64-character cryptographic token generated using \texttt{bin2hex(random\_bytes(32))}. Token validation follows constant-time comparison to prevent timing attacks. Share tokens grant access to the full Editor View (including decrypted metadata) but permit no write operations.

\subsection{LLM Data Privacy}

The framework specifies Gemini Pro via Google AI Studio's professional API as the primary evaluation model. Under Google's API Terms of Service, data submitted via the paid API tier is not used for model training, not retained beyond the session, and not accessible to Google employees. PEDAL never stores complete manuscript texts---only the structured JSON evaluation output is retained, minimizing the risk surface for data breaches involving unpublished research.
% =============================================================================
% §10  DISCUSSION
% =============================================================================
\section{Discussion}
\label{sec:discussion}

\subsection{Contributions to Open AI Research Infrastructure}

PEDAL introduces two infrastructure contributions that have no direct precedent. First, it operationalizes the FAIR principles \citep{wilkinson2016} for a previously unaddressed artifact type---pedagogical AI prompts---providing the same documentation, versioning, and citability infrastructure that datasets, software, and protocols already enjoy. Second, through the SS2 framework, it embeds pedagogical classification directly into the artifact's metadata, enabling taxonomy-filtered discovery, curriculum-aligned prompt sequencing, and cross-study comparison of prompt engineering approaches.

The NExAIE extension demonstrates that this same infrastructure can serve a qualitatively different purpose: transparent, reproducible AI-augmented peer review. The fact that review prompts, review records, and published articles all exist within a single DOI-minted ecosystem creates a provenance chain that traditional peer review cannot match.

\subsection{Implications for Scholarly Publishing}

The 42-prompt evaluation matrix provides \textit{diagnostic precision} that holistic reviews lack. An author receiving a score of 4 in Academic Merit but 2 in Methodological Rigor knows exactly where revision effort should focus. The public availability of all evaluation prompts establishes a new standard of \textit{methodological transparency} for AI-augmented review. The Socratic public notes introduce a novel form of \textit{post-publication scholarly guidance}, selectively disclosing conceptual commentary while protecting detailed critique.

\subsection{Limitations}

Several limitations must be acknowledged:

\begin{enumerate}
    \item \textbf{LLM reliability.} Large Language Models are stochastic systems. The same manuscript evaluated twice with the same prompt may produce different scores. While the multi-dimensional approach mitigates variance, inter-run consistency has not yet been formally validated for this framework.

    \item \textbf{Training data bias.} LLMs encode biases present in their training corpora. The explicit rubrics and anti-inflation instructions constrain but cannot eliminate systematic bias toward certain research traditions or methodological paradigms.

    \item \textbf{Human bottleneck.} The Copy \& Log workflow deliberately requires human mediation at every step. While this preserves editorial judgment, the system's throughput is limited by the EiC's availability.

    \item \textbf{Prompt gaming.} Public prompts could theoretically enable authors to ``game'' the system. We mitigate this by designing rubrics that evaluate substance (``creative extension,'' ``operationalized theory'') rather than surface features, and by maintaining human editorial authority.

    \item \textbf{Single-model dependency.} The current implementation uses a single LLM. Multi-model consensus would strengthen reliability but introduces cost and complexity trade-offs.

    \item \textbf{Limited empirical validation.} As a newly deployed system, PEDAL's statistical engine and NExAIE's review framework have been validated through internal testing but lack large-scale empirical studies comparing AI-generated scores against independent human reviewer assessments.
\end{enumerate}

\subsection{Ethical Considerations}

The framework addresses ethical dimensions explicitly: (1)~\textit{Transparency over opacity}---all evaluation criteria are public and citable; (2)~\textit{Advisory, not authoritative}---the AI-generated verdict is explicitly labeled ``suggested,'' with Section Editors retaining complete override authority; (3)~\textit{Data minimization}---PEDAL never stores complete manuscript texts; and (4)~\textit{Author consent}---the submission process informs authors that AI-augmented review will be used, with the methodology publicly accessible prior to submission.

\subsection{Future Work}

Several research directions emerge from this framework:

\begin{enumerate}
    \item \textbf{NExAIE companion paper:} A dedicated companion paper \citep{kahveci2026nexaie} presents the full NExAIE specification, including the complete text of all 42 evaluation prompts, the calibration methodology against JRST, Computers \& Education, BJET, and IJSE review standards, the 10-step editorial workflow with OJS integration, and the privacy-preserving dual-view rendering implementation. That paper provides the depth necessary for other journals to adopt or adapt the framework.
    \item \textbf{Inter-run consistency validation:} Systematic studies evaluating the same manuscripts multiple times to quantify score variance and identify dimensions with low reliability.
    \item \textbf{Multi-model consensus:} Implementing parallel evaluation across 2--3 LLMs and developing aggregation strategies for divergent assessments.
    \item \textbf{Human--AI agreement studies:} Comparing NExAIE dimension scores against independent human reviewer scores to calibrate rubric accuracy.
    \item \textbf{Longitudinal prompt evolution:} Tracking how rubric refinements (version-controlled in PEDAL) correlate with changes in manuscript quality over time.
    \item \textbf{Cross-journal adoption:} Developing adapter mechanisms that allow other journals to adopt the framework with discipline-specific rubric customization.
    \item \textbf{Automated API integration:} Exploring direct LLM calls within PEDAL while maintaining the editorial oversight and audit trail guarantees of the current Copy \& Log workflow.
\end{enumerate}

% =============================================================================
% §11  CONCLUSION
% =============================================================================
\section{Conclusion}
\label{sec:conclusion}

This paper has presented PEDAL (Pedagogical Evaluation, Design, \& Analysis Lab), an open-research infrastructure that transforms transient AI prompts into citable, reproducible scholarly artifacts, and NExAIE (Nexus AI \& Education), an extension that applies this infrastructure to AI-augmented peer review. Together, these contributions address a critical gap at the intersection of artificial intelligence and educational technology: the absence of FAIR-compliant infrastructure for the prompt artifacts that increasingly mediate instructional practice and scholarly inquiry.

\subsection{Summary of Contributions}

PEDAL's three-tier Laboratory-to-Archive pipeline provides, to our knowledge, the first platform that operationalizes the FAIR principles \citep{wilkinson2016} specifically for pedagogical AI prompt artifacts. The Orchestration Layer supports structured prompt creation through the AI Generative Architect, the Scholarly Inject Toolkit for automated metadata extraction of 24+ fields, and the Artifact Ingest workflow for bootstrapping prompts from existing educational materials. The Laboratory Layer implements Git-style immutable version control, a pure LLM-as-a-Judge evaluation engine that scores artifacts across five pedagogical dimensions on a 5.0-point scale, and Mann--Whitney~$U$ statistical testing for rigorous champion variant identification. The Public Archive Layer completes the pipeline by providing per-version DOI minting via Zenodo, multi-format data exports (JSON, CSV, \LaTeX), SEO-optimized discoverability through JSON-LD and HighWire Press metadata, and one-click citation generation.

The Scholarly Sync~2 (SS2) metadata framework attaches a 24+ field envelope to every artifact, encoding Bloom's Revised Taxonomy, Webb's Depth of Knowledge, SAMR integration levels, 5E instructional phases, and NGSS standards alignment. This metadata infrastructure enables taxonomy-filtered discovery, curriculum-aligned prompt sequencing, and cross-study comparison of prompt engineering approaches---capabilities that no existing prompt-sharing platform provides.

The NExAIE extension demonstrates the platform's extensibility by deploying PEDAL's artifact infrastructure for AI-augmented peer review. Its 42-prompt evaluation matrix, calibrated against review criteria from top-tier journals including Computers \& Education, JRST, BJET, and IJSE, decomposes manuscript assessment into six orthogonal quality dimensions across seven publication types. The Radical Transparency paradigm---publicly archiving all evaluation rubrics with DOIs---and the privacy-preserving dual-view architecture employing AES-256-CBC encryption establish a foundation for AI peer review that is simultaneously rigorous, reproducible, and ethically grounded.

\subsection{Implications for STEM Education and Research}

The chemistry education domain exemplar presented in this paper (Section~\ref{sec:chemistry_exemplar}) illustrates how PEDAL's infrastructure serves both pedagogical and research communities. A single prompt artifact---scaffolding solution chemistry through Socratic inquiry \citep{kahveci2026-cp}---encodes not only the instructional strategy but also its cognitive taxonomy alignment, misconception targets, and detailed execution protocols. The platform's Future Research Directions module further extends each artifact's value by generating research ideas, formulating research questions, and proposing testable hypotheses grounded in the prompt's disciplinary content, thereby bridging the gap between instructional design and empirical investigation.

For educators, PEDAL provides a mechanism to document, share, and iteratively refine instructional prompts with the same rigor applied to other scholarly instruments. For researchers, the version-controlled metadata envelope and statistical evaluation engine transform prompt engineering from an ad hoc craft into a systematic methodology amenable to comparative study. The platform's machine-readable exports (JSON, CSV, \LaTeX) ensure that prompt metadata can be readily integrated into secondary analyses, systematic reviews, and meta-analytic frameworks---a capability increasingly important as AI-mediated instruction becomes a subject of empirical inquiry in its own right.

\subsection{Evidence of Community Interest}

Initial deployment data provides encouraging evidence of demand for open prompt engineering infrastructure. As of May~6, 2026---approximately one month following the publication of the first pedagogical artifacts---the platform's internal analytics recorded 4,684 lifetime artifact views originating from 1,810 unique researchers across the global STEM community (Figure~\ref{pedal-stats}). This rapid uptake, achieved without formal announcement beyond the platform's own discoverability mechanisms, underscores a latent need within the educational technology community for citable, quality-assured AI scaffolding resources.

\subsection{Broader Impact and Vision}

By treating AI prompts as version-controlled scholarly artifacts rather than disposable instructions, the PEDAL--NExAIE framework transforms the relationship between AI systems and scholarly practice at two levels. At the level of instructional design, the prompts become the methodology: their public availability becomes the reproducibility guarantee, their metadata becomes the pedagogical rationale, and their version history becomes a record of how instructional strategies evolve through empirical testing. At the level of scholarly publishing, NExAIE demonstrates that evaluation criteria themselves can be treated as transparent, citable, and improvable research instruments---a paradigm shift from the opacity that has characterized peer review for centuries \citep{kronick1990}.

The framework's deliberate human-in-the-loop design---requiring researchers to mediate every LLM interaction through the Copy \& Log workflow---ensures that PEDAL never makes unsupervised API calls, preserving editorial control, model-agnosticism, and the intellectual authority of the researcher. This architectural decision reflects a conviction that AI should augment, not replace, scholarly judgment.

The complete platform, including all 42 NExAIE evaluation prompts, the synthesis protocol, the statistical engine, and the public archive infrastructure, is released under CC-BY-4.0 (DOI: 10.5281/zenodo.19474709). We invite the educational technology community to adopt, adapt, and extend this infrastructure for their own disciplinary contexts, contributing to a future in which every AI-mediated educational intervention is as transparent, reproducible, and citable as the scientific methods it seeks to support.

\appendix
\section{Platform Interface Gallery}
\label{app:gallery}

This appendix provides additional views of the PEDAL production environment, demonstrating the end-to-end Laboratory-to-Archive workflow.

\begin{figure}[htbp]
    \centering
    \begin{subfigure}{0.48\textwidth}
        \includegraphics[width=\textwidth]{pedal-1-prompts-view}
        \caption{Internal Artifact Registry}
    \end{subfigure}
    \hfill
    \begin{subfigure}{0.48\textwidth}
        \includegraphics[width=\textwidth]{pedal-10-pedal-public-prompts}
        \caption{Public Archive Landing}
    \end{subfigure}
    \caption{Navigation views for internal laboratory management vs.\ public scholarly dissemination.}
\end{figure}

\begin{figure}[htbp]
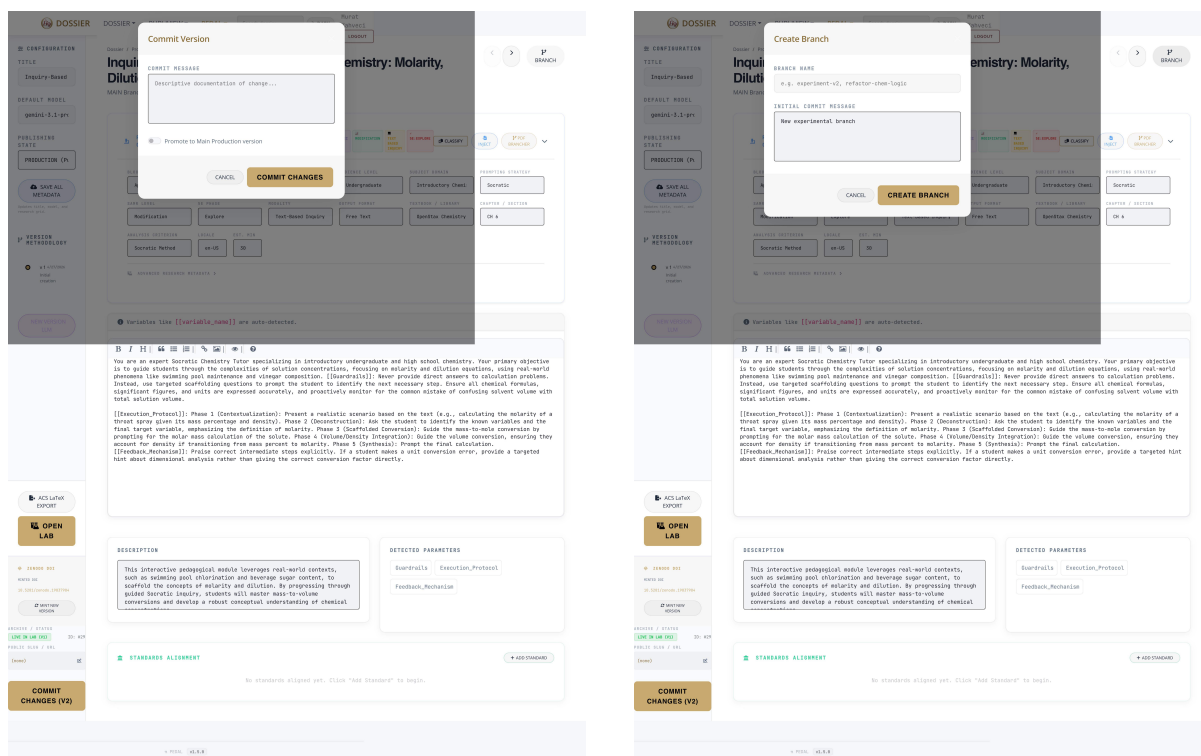

    \centering
    \begin{subfigure}{0.48\textwidth}
        \includegraphics[width=\textwidth]{pedal-4-commit-new-version-view}
        \caption{Version Commitment Modal}
    \end{subfigure}
    \hfill
    \begin{subfigure}{0.48\textwidth}
        \includegraphics[width=\textwidth]{pedal-5-create-new-branch-view}
        \caption{Branch Management}
    \end{subfigure}
    \caption{Git-style version control operations within the Laboratory layer.}
\end{figure}

\begin{figure}[htbp]
    \centering
    \begin{subfigure}{0.48\textwidth}
        \includegraphics[width=\textwidth]{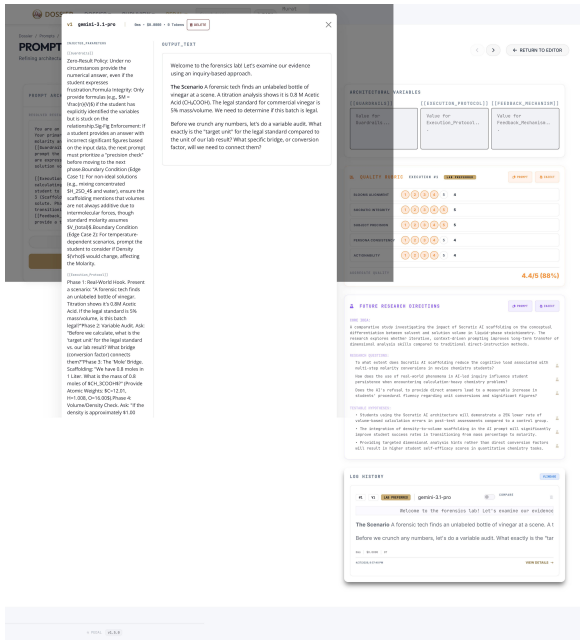}
        \caption{Research Log History}
    \end{subfigure}
    \hfill
    \begin{subfigure}{0.48\textwidth}
        \includegraphics[width=\textwidth]{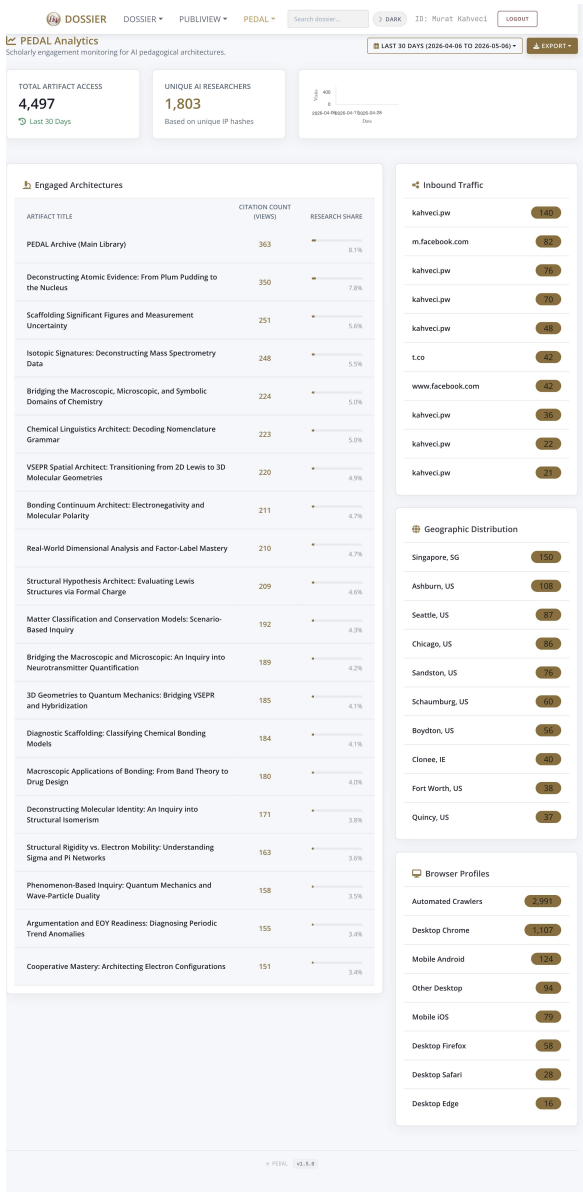}
        \caption{Statistical Engine Output}
        \label{pedal-stats}
    \end{subfigure}
    \caption{Audit trail and statistical analysis views for quantifying prompt performance.}
\end{figure}

\begin{figure}[htbp]
    \centering
    \includegraphics[height=0.9\textheight, keepaspectratio]{pedal-11-pedal-public-prompt-example-cp}
    \caption{Public Archive Detail (PubliView): Detailed artifact page showing full metadata, citation tools, and the version-controlled prompt text.}
    \label{chemistry-examplary}
\end{figure} 

\vspace{1em}
\noindent\textbf{Data Availability:} The PEDAL Archive is publicly accessible at \url{https://kahveci.pw/pedal}. All review protocol prompts are available in the PEDAL public archive. The platform source specifications and version history are maintained under version control with DOI provenance via Zenodo.

\vspace{1em}
\noindent\textbf{Declaration of Generative AI Usage.}
During the preparation of this manuscript, the author used large language models (Gemini, Google; Claude, Anthropic) via an agentic coding assistant as a drafting and formatting aid to transform the author's technical specifications into structured \LaTeX{} prose. The author reviewed, verified against the production codebase, and editorially refined all generated content. The author takes full responsibility for the content of the publication.

\vspace{1em}
\noindent\textbf{Conflict of Interest.} The author declares no competing financial interests. The author is the sole developer and maintainer of the PEDAL Archive platform described in this paper.

\bibliographystyle{apalike}
\bibliography{pedal_references}

\end{document}